\documentclass[twocolumn,prx,twocolumn,superscriptaddress]{revtex4}
\pdfoutput=1
\usepackage[utf8]{inputenc}
\usepackage{amsmath, amssymb,dsfont}
\usepackage{verbatim}
\usepackage[usenames,dvipsnames]{xcolor}
\usepackage{graphicx}
\usepackage{mathtools}
\usepackage{stmaryrd, ifthen}
\usepackage{array}
\usepackage{tabularx,float}
\usepackage[export]{adjustbox}
\usepackage{bm,wasysym}
\usepackage{enumitem}
\usepackage{calc}
\usepackage{etoolbox}
\mathtoolsset{showonlyrefs=false}
\usepackage{footnote}
\usepackage{dcolumn}
\usepackage{microtype}
\usepackage{booktabs}
\usepackage{epstopdf}

\usepackage{euscript}

\usepackage{amssymb,amsmath,amsfonts,bm,bbm} 

\usepackage{graphicx}
\usepackage{epstopdf}
\usepackage{times}
\usepackage{float}
\usepackage{lipsum}
\usepackage{color}
\usepackage{dcolumn}
\usepackage{bm}
\usepackage{subfigure}
\usepackage[colorlinks,linkcolor=blue,anchorcolor=blue,urlcolor=blue,citecolor=blue]{hyperref}
\usepackage{verbatim} 
\usepackage{braket}
\usepackage{bbold}

\providecommand{\theoremname}{Theorem}
\newcommand*{\myproofname}{Proof}

\newcommand*{\id}{\mathbbm{1}}
\newcommand*{\Tr}{\textrm{Tr}}

\usepackage{hyperref}
\hypersetup{colorlinks=true, linkcolor=blue, citecolor=red, urlcolor=blue  }

\begin{document}

\title{Unification of spatiotemporal quantum formalisms: mapping between process and pseudo-density matrices via multiple-time states}

\author{Xiangjing Liu}
\email{liuxj@mail.bnu.edu.cn}
\affiliation{Department of Physics, Southern University of Science and Technology, Shenzhen 518055, China}
\affiliation{Department of Physics, City University of Hong Kong, Kowloon, Hong Kong SAR }

\author{Zhian Jia}
\email{giannjia@foxmail.com}
\affiliation{Centre for Quantum Technologies, National University of Singapore, Singapore 117543, Singapore}
\affiliation{Department of Physics, National University of Singapore, Singapore 117543, Singapore}

\author{Yixian Qiu}
\affiliation{Centre for Quantum Technologies, National University of Singapore, Singapore 117543, Singapore}

\author{Fei Li}
\affiliation{Department of Physics, Southern University of Science and Technology, Shenzhen 518055, China}
\affiliation{Department of Physics, City University of Hong Kong, Kowloon, Hong Kong SAR }

\author{Oscar Dahlsten}
\email{oscar.dahlsten@cityu.edu.hk}
\affiliation{Department of Physics, City University of Hong Kong, Kowloon, Hong Kong SAR }
\affiliation{Shenzhen Institute for Quantum Science and Engineering, Southern University of Science and Technology, Shenzhen 518055, China}
\affiliation{Institute of Nanoscience and Applications, Southern University of Science and Technology, Shenzhen 518055, China}

\begin{abstract}  
	We consider the relation between three different approaches to defining quantum states across several times and locations: the pseudo-density matrix (PDM), the process matrix, and the multiple-time state approaches. Previous studies have shown that bipartite two-time states can reproduce the statistics of bipartite process matrices. Here, we show that the operational scenarios underlying two-time states can be represented as PDMs, and thereby construct a mapping from process matrices with measurements to PDMs. The existence of this mapping implies that PDMs can, like the process matrix, model processes with indefinite causal orders.  The results contribute to the unification of quantum models of spatiotemporal states. 
\end{abstract}

\maketitle


\section{Introduction}
Spatiotemporal quantum formalisms may be viewed as a way to overcome the disparity between the treatment of space and time in quantum mechanics. In spatial quantum processes, the state is typically described by a density operator, whereas in temporal quantum processes, the description involves input states and quantum channels~\cite{Nielsen2010}. The search for a formalism that treats temporal quantum processes as states, thereby eliminating or minimising the disparity, is thus a natural foundational line of inquiry~\cite{aharonov1964time,aharonov2009multiple,hardy2007towards,hardy2012operator,oreshkov2012quantum,fitzsimons2015quantum,horsman2017can,cotler2018superdensity,ohya1983note,Leifer2013toward,fullwood2022quantum,jia2023spatiotemporal}. Formalisms created to date, amongst other things (i) enable the investigation of temporal correlations in a similar manner to that of spatial correlations~\cite{leggett1985quantum,oreshkov2012quantum,fitzsimons2015quantum,liu2023quantum}, (ii) may provide a route towards a theory of quantum superposed space-times and quantum gravity~\cite{hardy2005probability,oreshkov2012quantum}, and (iii) provide a natural framework to investigate quantum causality~\cite{oreshkov2012quantum,fitzsimons2015quantum,liu2023quantum}.

Several concrete spatiotemporal formalisms have been proposed. These include consistent histories~\cite{griffiths1984consistent,gell1996quantum,omnes1992consistent,isham1994quantum,dowker1995properties,halliwell1994review}, pseudo density matrices (PDMs)~\cite{fitzsimons2015quantum,liu2023quantum}, quantum combs~\cite{giulio2009theoretical}, process matrices~\cite{oreshkov2012quantum,araujo2015witnessing},  multiple-time states~\cite{aharonov1964time,aharonov1991complete,aharonov2009multiple}, temporal compound states~\cite{ohya1983note}, Leifer-Spekkens causal states~\cite{Leifer2013toward}, superdensity operators~\cite{cotler2018superdensity}, symmetric bloom states~\cite{fullwood2022quantum,parzygnat2023timereversal} and doubled density operators~\cite{jia2023spatiotemporal}. Of particular relevance to the current work are (1) multiple-time states, which arise from treating quantum theory in a time-symmetric manner wherein both the initial and final states can be well-defined through pre and post-selection~\cite {aharonov1964time,aharonov1991complete,aharonov2009multiple}, (2) the PDM approach, which assigns an almost standard density matrix to temporally separated events and is therefore particularly convenient for extending correlation measures such as negativity into the time domain~\cite{fitzsimons2015quantum}, and (3) process matrices, which treat channels as matrices, in line with the Choi-Jamiolkowski isomorphism, and combine these with the initial state matrix into a matrix associated with the whole process~\cite{oreshkov2012quantum,araujo2015witnessing}.

Since the different approaches share many motivations and aims, it is natural that there is interest in understanding how concepts and statements can be transferred across different formalisms. In particular, the incomplete understanding of the relation between the approaches (1)-(3) hampers the ability to transfer insights gained in papers employing the different formalisms, such as how derivations concerning when the PDM has negativity relate to results concerning causality derived in the process matrix formalism. Certain recent results taken together give hope that a neat and general mapping between the formalisms can be established. Mappings from process matrices to PDMs were given in Ref.~\cite{zhang2020quantum} for the cases of one lab or one time though without a closed-form expression for the resulting PDM. Ref.~\cite{pusuluk2022witnessing} moreover used the multiple-time state formalism to define an intriguing new `single-time' variant of the PDM, but did not resolve how exactly the standard PDM definition relates to multiple-time states. Importantly for this paper, Ref.~\cite{silva2017connecting} connected the process matrix formalism with multiple-time states, showing that process matrices can be realised in standard quantum theory combined with pre and post-selection. Moreover, recent results gave closed-form expressions for PDMs~\cite{zhao2018geometry,horsman2017can, liu2023quantum} making it more tractable to express a connection between the three approaches neatly and in generality.  

We accordingly here aim to clarify the relationship between the PDM and the other two approaches. Leveraging the closed-form expression for the PDM of Ref.~\cite{liu2023quantum}, we derive a map between multiple-time states and PDMs which preserves the statistics. This then, via the mapping in Ref.~\cite{silva2017connecting}, leads to a mapping from process matrices with a positive operator-valued measure (POVM) to PDMs. The results enable an increased transferability of insights between papers employing the different approaches. In particular, the existence of the mapping implies that process matrix scenarios with indefinite causality can also be modelled in the PDM formalism.

The rest of the paper is organized as follows: In Sec.~\ref{sec:pre}, we provide a brief review of the PDM, two-time state, and process matrix formalisms. Sec.~\ref{sec:BornRule} generalises the Born rule for PDMs to cases with probabilistic maps, which is then used in Sec.~\ref{sec:PDM2TimeState} to establish the connection between PDMs and two-time states. The next connection, between PDMs and process matrices, is then established in Sec.~\ref{sec:PDMprocess} with two-time states acting as a bridge between these two formalisms.  Finally, in Sec.\ref{sec:conclusion}, we provide concluding remarks and discuss the future prospects of our findings.

\section{ Preliminaries}
\label{sec:pre}
In this section, we briefly review three main formalisms of the temporal quantum process that we are concerned with in this work: the pseudo-density matrix formalism; the two-time state formalism, and the process matrix formalism.

\subsection{Pseudo-density matrix formalism}
The PDM is a generalized framework designed to treat space and time on an equal footing~\cite{fitzsimons2015quantum}. 
The PDM formalism entails defining events by means of measurements conducted in space-time and relies on the correlations between the outcomes of these measurements. 
Therefore it treats temporal correlations the same way it treats spatial correlations, integrating them into a unified framework.
The PDM has been employed to explore various aspects of temporal quantum processes~\cite{liu2023quantum,zhang2020quantum,zhang2020different,Pisarczyk2019causal,marletto2019theoretical,jia2023quantum,zhao2018geometry,marletto2021temporal}, including quantum causal inference~\cite{liu2023quantum,liu2023inferring}, quantum communication~\cite{Pisarczyk2019causal}, temporal quantum teleportation~\cite{marletto2021temporal}, and compatibility of quantum processes~\cite{jia2023quantum}, among others.

The PDM formalism can be viewed as a generalization of the standard quantum $n$-qubit density matrix to the case of multiple times. The PDM is defined as~\cite{fitzsimons2015quantum}
\begin{align}\label{eq: defPDM}
	R_{1\cdots m}&=\frac{1}{2^{mn}} \sum^{4^n-1}_{i_1=0}\cdots\sum^{4^n-1}_{i_m=0}  \langle  \{ \tilde{\sigma}_{i_{\alpha}} \}^m_{\alpha=1} \rangle   \bigotimes^m_{\alpha=1} \tilde{\sigma}_{i_{\alpha}} \nonumber\\
	&\in \mathcal{L}( \mathcal{H}^{t_1}\otimes\cdots\otimes \mathcal{H}^{t_m })  ,
\end{align}
where $  \tilde{\sigma}_{i_\alpha}\in \{ \sigma_0,\sigma_1,\sigma_2, \sigma_3\}^{\otimes n} $ is an $n$-qubit Pauli matrix at time $t_\alpha$. $\tilde{\sigma}_{i_\alpha}$ is extended to an observable associated with $m$ times, $\bigotimes^m_{\alpha=1} \tilde{\sigma}_{i_{\alpha}}$ that has expectation value  $ \langle  \{ \tilde{\sigma}_{i_\alpha} \}^m_{\alpha=1} \rangle $.  By tracing out the Hilbert spaces for all time points except one, $t_{\alpha'}$, we recover the conventional quantum density matrix, i.e.,\ $\rho_{\alpha'}=\Tr_{\alpha \neq \alpha'}R_{1\cdots m}$. The PDM has unit trace and is Hermitian. However, it may have negative eigenvalues.

The negative eigenvalues of the PDM are found in a measure of temporal entanglement called a causal monotone $f(R)$~\cite{fitzsimons2015quantum}. Similar to entanglement monotones~\cite{vidal2000entanglement}, $f(R)$ is generally expected to fulfill the criteria: I) $f(R) \geq 0 $; II) $f(R)$ is invariant under local change of basis; III)  $f(R)$ is non-increasing under local operations; and IV) $ \sum_i p_i f(R_i) \geq f(\sum p_i R_i)$.
The criteria are satisfied by~\cite{fitzsimons2015quantum}:
\begin{align}\label{eq: causmono}
	f(R) :=||R||_{\rm tr}-1 = \Tr \sqrt{ R^\dag R }-1.
\end{align} 
If $R$ has negativity, $f(R)>0$.We will refer to $f(R)$ as the PDM negativity in this work. The presence of negative eigenvalues indicates the existence of temporal correlations. This is due to the fact that the standard density matrix at a single time is positive semi-definite and cannot account for the negativity.

A closed-form for the PDM with $n$ qubits at $m$ times has recently been discovered ~\cite{liu2023quantum}.
The closed form is obtained by utilizing the Choi-Jamio{\l}kowski (CJ) matrix of a completely positive trace-preserving (CPTP) map or quantum channel $ \mathcal{M}_{i+1|i}: \mathcal{L}(\mathcal{H}^{t_i}) \rightarrow  \mathcal{L}(\mathcal{H}^{t_{i+1}})  $ in the time interval $[t_i,t_{i+1}]$. The CJ matrix is defined as follows~\cite{choi1975completely,jamiolkowski1972linear},
\begin{equation} \label{eq: CHOI}
	M_{i,i+1}:= \sum^{2^n-1}_{m,n=0}\ket{m}\bra{n}^T \otimes  \mathcal{M}_{i+1|i} \left( \ket{m}\bra{n}  \right),
\end{equation}
where the superscript $T$ denotes the transpose. In accordance with  Eq.~\eqref{eq: defPDM},  the key to obtaining a closed form is to derive an analytic form of the expectation values $ \langle  \{ \tilde{\sigma}_{i_\alpha} \}^m_{\alpha=1} \rangle $. The measurement scheme at each time $t_\alpha$ for determining the expectation values $ \langle  \{ \tilde{\sigma}_{i_\alpha} \}^m_{\alpha=1} \rangle $ is chosen to be the coarse-grained projectors,
\begin{align}\label{Eq:coarsegrainedM}
	\Bigl\{ P^{i_\alpha}_+= \frac{ \mathbbm{1}  + \tilde{\sigma}_{i_\alpha}}{2} , 
	P^{i_\alpha}_-= \frac{\mathbbm{1}  - \tilde{ \sigma}_{i_\alpha}}{2} \Bigl\}, 
\end{align}
where $\alpha$ labels the time of the measurement. Denote $\rho_1$ by the state of interest at time $t_1$. Following the above, the $n$-qubit PDM across $m$ times is given by the following iterative expression~\cite{liu2023quantum}
\begin{align} \label{eq: PDMclosedForm}
	R_{1\cdots m}= \frac{1}{2} (R_{1\cdots m-1} M_{m-1,m}+  M_{m-1,m} R_{12\cdots m-1}),
\end{align}
with the initial condition $R_{12}= \frac{1}{2}( \rho \, M_{12} +M_{12} \, \rho ) $ where $\rho:= \rho_1 \otimes \id_2 $.  Note that although its derivation concerns qubits, the closed-form PDM can describe a quantum system of any finite dimensionality since one can embed such a system into a state space of qubits and confine its evolution within an appropriate subspace. Nevertheless, we consider quantum systems made of qubits throughout this work to simplify our considerations. 
 It is worth mentioning that even though the derivation of the closed-form of PDM is operational in nature, some references perceive it as a physical object~\cite{horsman2017can,fullwood2022quantum,parzygnat2023timereversal}.

\subsection{The two-time state formalism}
The two-time formalism offers a comprehensive framework for analyzing quantum systems that involve two distinct points in time~\cite{aharonov1964time, aharonov1991complete}. These states incorporate information about both the pre-selected and post-selected states, providing a comprehensive description of the system. The study of two-time states gained renewed interest with the discovery of weak measurements \cite{Aharonov1988how}. Subsequently, the formalism was extended to the case of multiple times \cite{aharonov2009multiple}, allowing for a more comprehensive understanding of temporal quantum behavior.

In the following, we provide a concise overview of the two-time state formalism. Consider the following experimental process.  Alice prepares an initial state $\ket{\psi}$ at time $t_1$, then performs a set of operators $\{\hat{E}_a\}$ on the state. After that, Alice, at time $t_2$, performs a projective measurement with $ \ket{\phi} \bra{\phi} $ being one of the projectors. The results of the whole experiment are kept when Alice observes the state $\ket{\phi}$ at time $t_2$, otherwise, discarded.   The operator $\hat{E}_a$ is called a Kraus operator and satisfies the completeness relation $\sum_{a} \hat{E}^\dag_a \hat{E}_a =\id $. Alice has the ability to record the outcome $a$ of $\hat{E}_a$.  The set $\{\hat{F}_a:=  \hat{E}^\dag_a \hat{E}_a  \}$ forms a POVM. $\hat{E}_a$ is also associated with a completely positive trace-non-increasing (CP) map $\mathcal{M}_a$, which acts on a density matrix $\rho$ as $\mathcal{M}_a(\rho) = \hat{E}_a \rho \hat{E}_a^\dag$.   In this process, the probability of Alice implementing the CP map $\mathcal{M}_a$ given the pre-selected state $\ket{\psi}$ and post-selected state $\ket{\phi}$ is given by
\begin{align} \label{eq: SimpleStatProb}
	P(a) = \frac{ | \bra{\phi} \hat{E}_a \ket{ \psi }  |^2  }{ \sum_{a'} | \bra{\phi} \hat{E}_{a'} \ket{ \psi }  |^2   },
\end{align}
which is also known as the Aharonov-Bergman-Lebowitz (ABL) rule \cite{aharonov1964time}.

We now describe the above process in the two-time state formalism. We adopt notations from Ref.~\cite{silva2017connecting}.  The Hilbert space of Alice across two times $t_1, t_2$ is represented by $\mathcal{H}_{A_2}  \otimes \mathcal{H}^{A_1}$. The Hilbert space $\mathcal{H}^{A_1}$ consisting of ket vectors represents the pre-selected state space where states evolve forward in time. In contrast, $ \mathcal{H}_{A_2}$ consisting of bra vectors represents post-selected state space where states evolve backward in time.  Then, the corresponding two-time state and Kraus operator $\hat{E}_a$ can be expressed as~\cite{Silva2014pre,silva2017connecting}
\begin{align}
	\Psi &= {}_{A_2}\!\bra{\phi} \otimes  \ket{\psi}^{A_1}\;\; \in \mathcal{H}_{A_2} \otimes \mathcal{H}^{A_1}, \nonumber \\
	E_a &= \sum_{kl} \beta_{a,kl} \ket{k}^{A_2} \otimes  {}_{A_1}\!\bra{l} \;\;  \in \mathcal{H}^{A_2} \otimes \mathcal{H}_{A_1}.
\end{align}
In a similar fashion to the density matrix in standard quantum mechanics, the \emph{density vectors} for the above two-time state $\Psi$ and the Kraus operator $E_a$ are defined as \cite{Silva2014pre,silva2017connecting}:
\begin{align} \label{eq: DensityVector}
	\Psi \otimes \Psi^\dagger  =& 
	{}_{ A_2}\!\bra{ \phi  } \otimes \ket{ \psi  }^{ A_1} \otimes {}_{ A_1^\dagger}\!\bra{ \psi  } \otimes \ket{ \phi }^{A_2^\dagger}  \nonumber\\
	&\in  \mathcal{H}^{\mathcal{A}_1 } \otimes  \mathcal{H}_{\mathcal{A}_2},
	\nonumber\\
	J_a  =& E^a \otimes E^\dag_a \in    \mathcal{H}_{\mathcal{A}_1 } \otimes  \mathcal{H}^{\mathcal{A}_2},
\end{align}
where the notation $ \mathcal{H}^{ \mathcal{A} } := \mathcal{H}^{ A } \otimes \mathcal{H}_{ A^\dagger}   $ and $ \mathcal{H}_{\mathcal{A}} := \mathcal{H}_{ A } \otimes \mathcal{H}^{ A^\dagger}   $ is used. 
One immediately sees that the state and the measurement are both defined in the Hilbert spaces by those notations and are in that sense placed on an equal footing.  To contract bra and ket vectors, the  $\bullet$ operation is introduced in the two-time formalism and is defined via 
\begin{align}
	{}_{A} \bra{\phi} \bullet \ket{ i }^{A} = \ket{ i }^{A} \bullet {}_A \bra{ \phi }  := \braket{ \phi }{i}.
\end{align}
A direct calculation shows that
\begin{align} 
	| \Psi \bullet E_a |^2  = (     \Psi \otimes \Psi^\dagger) \bullet J_a = | \bra{\phi} \hat{E}_a \ket{ \psi }  |^2 .   \nonumber
\end{align}
The ABL rule of Alice's operation $\mathcal{M}_a$ by Eq.~\eqref{eq: SimpleStatProb}  now takes on an equivalent but different form,
\begin{align} \label{eq: Simp2TimeProb}
	P_T(a)= \frac{ | \Psi \bullet E_a |^2 }{ \sum_{a'} |  \Psi \bullet E_{a'}   |^2 } = \frac{ (      \Psi \otimes \Psi^\dagger) \bullet J_a }{ \sum_{a'}  (  \Psi \otimes \Psi^\dagger ) \bullet J_{a'}  } ,
\end{align}
where the subscript $T$ denotes probabilities in the multiple-time formalism.
The generalization of the above expression to any pure two-time states is straightforward. A general pure two-time state takes the form
\begin{align}
	\Psi &= \sum_{ij} \alpha'_{ij} \, {}_{A_2}\!\bra{ i } \otimes  \ket{j }^{A_1},
\end{align}
where $\alpha'_{ij}$ can be any complex number,  and its corresponding pure density vector is 
\begin{align}
	&\Psi \otimes \Psi ^\dag\nonumber \\
	=& \sum_{i,j,m,n} \alpha'_{ij} \alpha^{'*}_{mn} \, {}_{A_2}\!\bra{ i } \otimes  \ket{j }^{A_1} \otimes  {}_{A_1^\dag }\!\bra{ n } \otimes  \ket{ m  }^{A_2^\dag } .
\end{align}
Such states can be physically realized via post-selecting on maximally entangled states and tracing out, as will be explained later. 
The probability of Alice's outcome is given by Eq.~\eqref{eq: Simp2TimeProb} also for that general pure state.

Probabilistic combinations of states may also be modeled in this manner. 
An {\em ensemble} of pure two-time states $\{ p_r, \Psi_r:= \sum_{ij} \alpha'_{r, ij} \, {}_{A_2}\!\bra{ i } \otimes  \ket{j }^{A_1}  \}$ is a probabilistic combination of the pure density vectors \cite{Silva2014pre},
\begin{align}
	\eta= \sum_{r} p_r \Psi_r \otimes \Psi_r^\dag .
\end{align}
The statistics of Alice's operation in the ensemble of two-time states are then given by
\begin{align} \label{eq: Prob2TimeEnsemb}
	P_T(a)=  \frac{ \sum_r p_r    (   \Psi_r \otimes \Psi_r^\dagger) \bullet J_a }{ \sum_{a',r}  p_r (  \Psi_{r} \otimes \Psi_{r}^\dagger ) \bullet J_{a'}  } = \frac{ \eta \bullet J_a }{ \sum_{a'} \eta \bullet J_{a'}   }.
\end{align}

 For coarse-grained CP maps $\mathcal{M}_a $ with action $ \mathcal{M}_a(\rho):= \sum_\mu \hat{E}^\mu_a \rho \hat{E}^\mu_a $, the probabilities in the above three scenarios have the same expression as Eqs.~\eqref{eq: SimpleStatProb}, \eqref{eq: Simp2TimeProb} and \eqref{eq: Prob2TimeEnsemb}, respectively~\cite{silva2017connecting}.

\subsection{The process matrix formalism}
The quantum process matrix formalism was introduced to describe `indefinite causal orders'~\cite{oreshkov2012quantum} in which there is a quantum superposition of causal orders of events. In this subsection, we will briefly describe the process matrix formalism for the case of two parties, Alice and Bob.

The bipartite process matrix formalism is formulated in a signaling game~\cite{oreshkov2012quantum}. Two experimenters Alice and Bob, reside in separate laboratories and they are allowed to send signals (quantum system) to each other in a game. Within their local laboratories, the causal order is definite, however, to achieve a better strategy for the game, the global causal structure between the two laboratories is not assumed to have a pre-defined order.  More formally, during each run of the game, each party receives a quantum system and performs quantum operations on it, after which the quantum system is then sent out of the laboratory. The laboratories are only opened to allow the system to come in and out, remaining closed during the period in which operations are carried out. The above can be modeled by a so-called process matrix.  The bipartite process matrix specifies how the two labs are connected and provides quantum systems to each lab.

Mathematically, the bipartite process matrix $ W^{A_1 A_2 B_1 B_2} $ is an object in $\mathcal{L}(\mathcal{H}^{A_1} \otimes \mathcal{H}^{A_2}\otimes \mathcal{H}^{B_1}\otimes \mathcal{H}^{B_2})$ where  $\mathcal{H}^{X_1}$ and $\mathcal{H}^{X_2}$ with $X=A, B$ denote the input and output Hilbert spaces for Alice and Bob respectively. Suppose that the complete sets of quantum operations for the two laboratories are chosen as $\{ \mathcal{M}_a^A \}$ and $\{ \mathcal{M}_b^B \}$ respectively.  The generalized Born rule for the process matrix is of the form
\begin{equation}\label{eq: BornRulePM}
	P_W(a,b)=\Tr [W^{A_1A_2B_1B_2}(M^a \otimes M^b)],
\end{equation}
where the subscript $W$ denotes the probabilities in the process matrix formalism and $M^a$ and $M^b$ are the CJ matrices for $\mathcal{M}_a^A$ and $\mathcal{M}_b^B$, respectively. A valid process matrix $W$  is required to satisfy two conditions, (i) it is positive semidefinite such that probabilities in Eq.~\eqref{eq: BornRulePM} are non-negative for all quantum operations; (ii) the probability distribution $P_W(a,b)$ generated by $W$ for any $ \mathcal{M}^A_a, \mathcal{M}^B_b $is normalized, i.e., $\sum_{ab} P_W(a,b)=1 $. For a more detailed introduction please refer to Refs.~\cite{oreshkov2012quantum,araujo2015witnessing}.

The process matrix formalism is a natural tool to deal with causal non-separability or indefinite causality. A causal structure consists of a collection of event locations accompanied by a partial order $\preceq$, which establishes the potential causal connections between events situated at these positions. In the bipartite scenarios, $A \preceq B$ denotes `$A$ is in the causal past of $B$'.  If $A$ is not in the causal past of B, the notation $A \npreceq B$ is used and the process matrices that describe this kind of process are denoted by $W^{A \npreceq B}$. A scenario in which $B \npreceq A$ with a probability of $ q \in [0,1]$, 
and $A \npreceq B$ with a probability of $ 1-q $, is modelled by a process matrix of the form,
\begin{align}
	\label{eq:causalseparability}
	W^{A_1 A_2 B_1 B_2} = q W^{B \npreceq A}+(1-q) W^{ A \npreceq B}.
\end{align}
The processes of this kind are called causally separable. Furthermore, processes that do not allow such a decomposition to exist are referred to as causally non-separable, or equivalently, exhibiting indefinite causality. A simple but insightful example of causally non-separable processes is the quantum switch~\cite{giulio2009theoretical}. It is an operation that superposes the orders of two events, and the corresponding process matrix and the proof of its causal non-separability are provided in Ref.~\cite{araujo2015witnessing}.

\section{Probability of a temporal sequence of outcomes given a PDM}
\label{sec:BornRule}

\begin{figure}[h]
	\centering
	\includegraphics[scale=0.45]{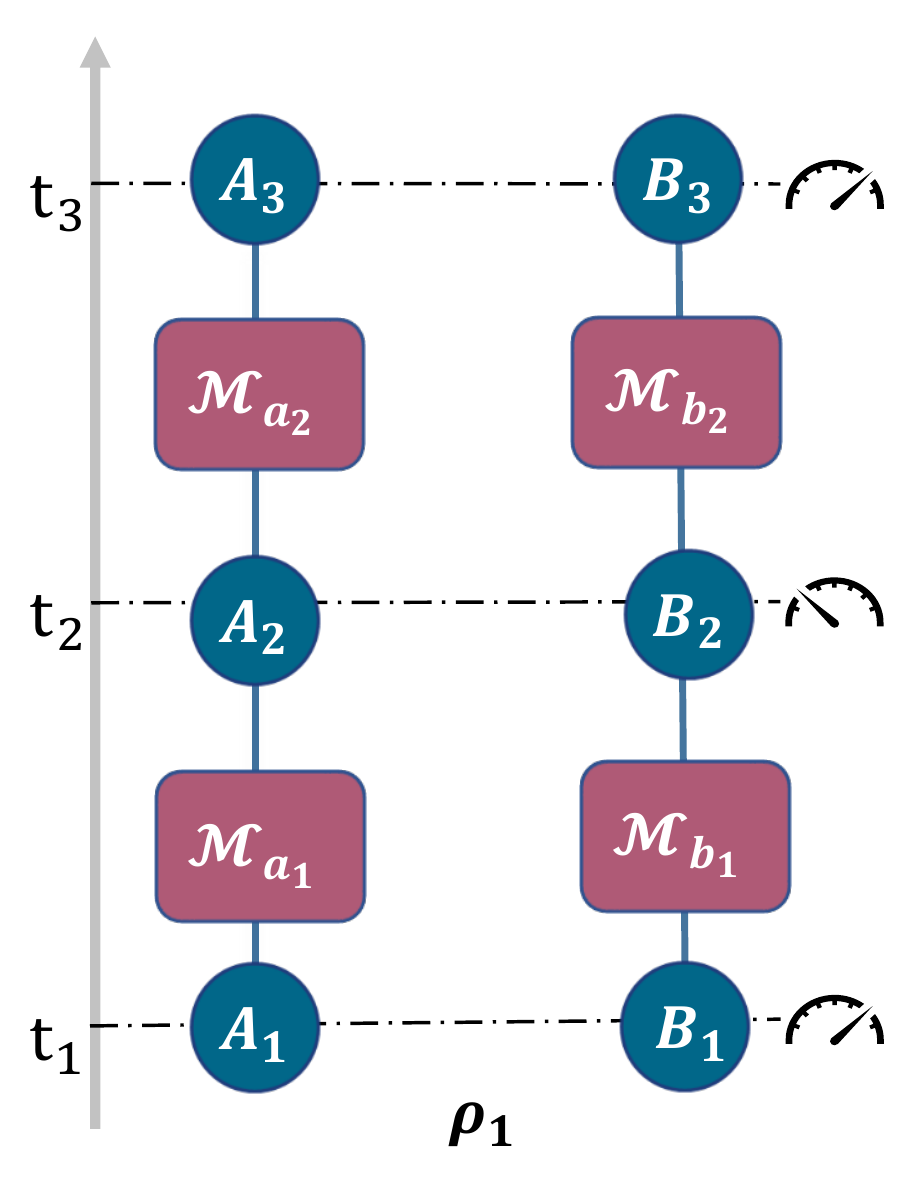}
	\caption{ \textbf{ Scheme for constructing the PDM for two parties across three times.} Coarse-grained measurements described by Eq.~\eqref{Eq:coarsegrainedM} are taken on the compound system $AB$ at three time points: $t_1, t_2,$ and $t_3$. Each party performs quantum operations on their respective part of the system during the two time intervals.}
	\label{fig:BornRule}
\end{figure}

In anticipation of mapping multiple-time states and process matrices to corresponding PDMs, we now derive a probability rule for PDMs. The rule extracts probabilities encoded in PDMs. It also applies to probabilities of maps that occur between the spatiotemporal tomographic measurements. 

Let us first consider probabilities in a two-time PDM $R_{12}$. The construction of a two-time PDM consists of the measurement results of correlators $ \langle \{ \tilde{\sigma}_{i_1},   \tilde{\sigma}_{j_2}\} \rangle $ and the corresponding Pauli operators, $i_1,j_2=0,1,...,2^n-1$. Each correlator $\langle \{ \tilde{\sigma}_{i_1},   \tilde{\sigma}_{j_2}\} \rangle$ is measured in separate repeated experiments. When  $\tilde{\sigma}_{i_1}$  and $ \tilde{\sigma}_{j_2}$ are not the identity operator the PDM only encodes the corresponding temporally coarse-grained probabilities, $ P( P^{i_1} P^{j_2} =1 ):=P(P^{i_1}=1,   P^{j_2} = 1)+P(P^{i_1} =-1,   P^{j_2} = -1) $ and $ P( P^{i_1}   P^{j_2} =-1):=P( P^{i_1}=-1,   P^{j_2} = 1)+P(P^{i_1}=1,  P^{j_2} = -1) $. The coarse-grained probabilities can be extracted from the PDM through the following two equations: 
 \begin{align} \label{eq:temporalCS}
    & P( P^{i_1}   P^{j_2} = 1)-P(P^{i_1}   P^{j_2} = - 1) = \Tr ( \tilde{\sigma}_{i_1} \otimes  \tilde{\sigma}_{j_2}\, R_{12}), \nonumber\\
    &  P( P^{i_1}   P^{j_2} = 1)+P(P^{i_1}   P^{j_2} = - 1) = 1.
 \end{align}
 It is thus clear that only temporally coarse-grained probabilities are needed to construct a PDM. This point also justifies the measurement scheme proposed in Refs.~\cite{miquel2002interpretation, souza2011scattering}, in which an ancilla is introduced to measure those temporally coarse-grained probabilities instead of directly measuring the system of interest.
 
 One crucial difference in probabilities between spatially and temporally distributed quantum systems is that they obey different rules. In the spatial case, the PDM returns to the normal density matrix, and all the spatially coarse-grained probabilities defined in Eq.~\eqref{Eq:coarsegrainedM} together render the complete information of any quantum state, such that other fine-grained probabilities can be extracted via the standard Born rule in quantum mechanics.
  However, this is not the case when dealing with temporally distributed quantum systems, since measurements made at earlier times influence measurement outcomes at later times.

There may also be a party `Alice' that is located in between those Pauli observables made at $t_1,t_2$.  Alice takes input from time $t_1$,  performs a quantum operation and then gives an output at time $t_2$. Generally, a quantum operation can be realized by performing a joint unitary transformation on both the input system and an ancilla,  then conducting a projective measurement on one part of the resulting joint system, leaving the other part as an output. Each outcome $a$ of the projective measurement induces a CP map $\mathcal{M}_a$ on the system. The set of CP maps $ \{ \mathcal{M}_a \}$ corresponding to all possible projective outcomes sum up to a CPTP map, i.e., $ \sum_a  \mathcal{M}_a:= \mathcal{M} $ is a CPTP map.
The CJ matrix for the map  $ \mathcal{M} $ is expressed by
\begin{align}
	M=\sum_a M^a := \sum_a \sum_{m,n} \ket{ n } \bra{ m } \otimes   \mathcal{M}_a ( \ket{ m } \bra{n}  ) .
\end{align}
Note that the time label for the CJ matrices is omitted since there is no confusion and instead the subscript $a$ here denotes Alice's outcome. The corresponding two-time PDM for the measurement scheme of Eq.\eqref{Eq:coarsegrainedM}, according to Eq.~\eqref{eq: PDMclosedForm}, is given by 
\begin{align}\label{eq:ProbCP}
	R_{12}=& \frac{1}{2}( M \, \rho + \rho \, M)=  \sum_a \frac{1 }{2} (M^a \ \rho + \rho \ M^a ) \nonumber\\
	:=&  \sum_a R_{12}^a .
\end{align}

Combining Eq.~\eqref{eq:temporalCS} with Eq.~\eqref{eq:ProbCP}, the joint probability of $  \tilde{\sigma}_{i_1}   \tilde{\sigma}_{j_2} $ ( $\tilde{\sigma}_{i_1}$ and $\tilde{\sigma}_{j_2}$ are not the identity operator)  and Alice recording the outcome $a$ 
 is given by 
\begin{align} \label{eq: ProbTemporalCG}
    	P_R( P^{i_1} P^{j_2} =\pm 1 ,a) =  \frac{1}{2}( 1 \pm \Tr (  \tilde{\sigma}_{i_1} \otimes \tilde{\sigma}_{j_2} R_{12}^a ) ).
\end{align}
The probability rule for measuring an $n$-qubit Pauli operator at one time while doing nothing at another time is then given by
\begin{align}\label{eq:ProbStandBorn}
   & P_R ( \Pi_{i_1},a)=  \Tr( \Pi_{i_1} \otimes \id \, R^a_{12}), \nonumber\\
   &     P_R ( \Pi_{j_2},a)=  \Tr(  \id \otimes \Pi_{j_2} \, R^a_{12}),
\end{align}
where $\Pi$ denotes the eigen-projector of an $n$-qubit Pauli matrix. The probability of doing nothing at both events  and Alice recording the outcome $a$ is given by
\begin{align} \label{eq: BornRulePDM}
	P_R(a)=\Tr  \mathcal{M}_a(\rho_1)=  \Tr R_{12}^a ,
\end{align}
where the property of the PDM $\Tr_1 R_{12} = \mathcal{M}(\rho_1)$ is used in the second equality~\cite{liu2023quantum}.  Eq.~\eqref{eq: BornRulePDM}
will play a central role in reproducing statistics in a process matrix.

The generalization to multi-party PDM across multiple times is straightforward.  For simplicity, we here only consider two parties Alice and Bob across three times $t_1,t_2,t_3$, as shown in FIG.~\ref{fig:BornRule}. During the interval $[t_1,t_2]$, Alice and Bob perform the CP maps $\mathcal{M}^A_{a_1} $ and $  \mathcal{M}_{b_1}^B $ on their part of the system.  Similarly, in the interval $[t_2,t_3]$, quantum operations  $\mathcal{M}^A_{a_2} $ and $  \mathcal{M}_{b_2}^B$ are implemented.  Denote $M^{a_1b_1}, M^{a_2b_2}$ by the CJ matrices of  $\mathcal{M}^A_{a_1} \otimes \mathcal{M}_{b_1}^B $ and $\mathcal{M}^A_{a_2} \otimes \mathcal{M}_{b_2}^B $, respectively. The three-time PDM corresponding to FIG.~\ref{fig:BornRule}, is written as 
\begin{align}
	R^{a_1b_1a_2b_2}_{123}=  \frac{1 }{2} ( M^{a_2b_2} \ R^{a_1b_1}_{12} + R^{a_1b_1}_{12} \  M^{a_2b_2} ) ,
\end{align}
where $    R^{a_1b_1}_{12}=  \frac{1 }{2} ( M^{a_1b_1} \ \rho + \rho \  M^{a_1b_1} )$. 
The probability for doing nothing at $t_1, t_2$ and Alice recording the outcomes $ a_1,b_1,a_2,b_2$ is given by 
\begin{align}
	P_R(a_1,b_1,a_2,b_2)=  \Tr_{123} R^{a_1b_1a_2b_2}_{123}.
\end{align}
The generalization of Eqs.~\eqref{eq: ProbTemporalCG} and~\eqref{eq:ProbStandBorn} is omitted since it is not needed to reproduce the statistics in the process matrix and multi-time state.

To conclude this section, the PDM returns to the conventional density matrix at each instance of time and thus it contains all the spatial correlations. However, temporally, the PDM encodes only the  coarse-grained probabilities $ P( P^{i_1}   P^{j_2} = \pm 1)$, without access to more fine-grained probabilities, such as $ P( P^{i_1}=1,  P^{j_2} = 1)$, according to the probability rule in a two-time PDM. This restriction in the temporal direction suggests the expressibility of the PDM representation is more restricted than that of double-Hilbert space representations for operational events, wherein there is one space for the input and another for the output.

\section{Mapping two-time states to PDMs }
\label{sec:PDM2TimeState}

In this section, we construct the mapping between the PDM and the two-time state formalisms. Specifically, each physical realization of a two-time state with dynamic schemes can be modeled by a PDM, exhibiting the same statistics for quantum operations. We assume that the physical realization only involves qubits such that the closed-form PDM in Ref.~\cite{liu2023quantum} can be used directly, as mentioned in Sec.~\ref{sec:pre} A.

\begin{figure*}
	\centering
	\includegraphics[scale=0.45]{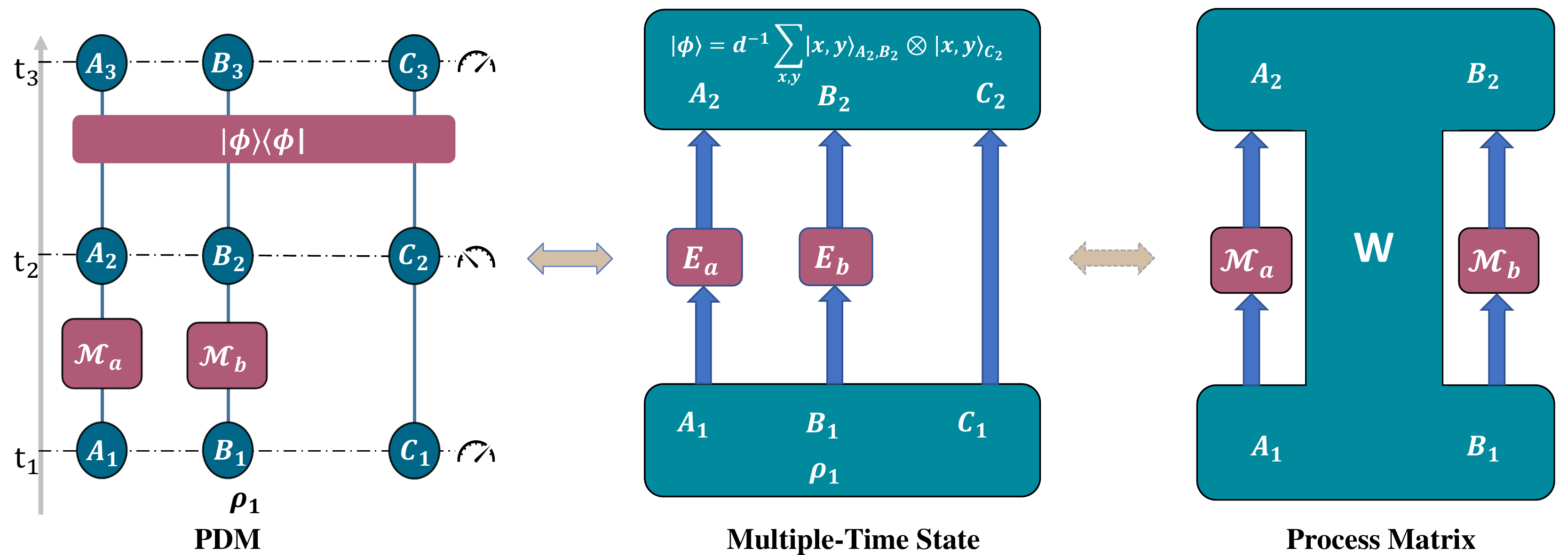}
	\caption{\textbf{Mappings between PDM, multiple-time state, and process matrix}. The rightmost figure depicts a bipartite process matrix $W$ with a POVM as in Eq.~\eqref{eq: BiProcessMatrix}. The middle figure is one physical realization of the two-time state $\eta$ in Eq.\eqref{eq: Bi2TimeStat}  with an operator scheme in the middle. The leftmost figure depicts the PDM formulation of the physical realization of $\eta$ with a operator scheme. We give a mapping from multiple-time states to PDMs, which together with the known mapping from process matrices to multiple-time states yields a mapping from given process matrices with a POVM to corresponding PDMs.
	}
	\label{fig:correspondence}
\end{figure*}

\subsection{ PDM for a simple two-time state}

We begin with the simplest two-time state $ {}_{A_2} \bra{i} \otimes \ket{j}^{A_1}$
and the dynamics given by  the set of Kraus operators $ \{ \hat{E}_a \} $. 
The physical realization is as follows.  Alice prepares a state $\ket{j }$ at time $t_1$ and selects the state $\ket{i}$ at time $t_2$.  In the intermediate time, Alice performs a detailed measurement, described by $ \{\hat{E}_a \} $. 
Each Kraus operator $\hat{E}_a$ corresponds to a CP maps $ \{ \mathcal{M}_a \}$ with the action $\mathcal{M}_a(\rho) =  \hat{E}_a \rho \hat{E}^\dag_a  $. 

We now turn to model the physical process described above using the PDM.
The dynamics of the system consist of two processes, the CP map $ \mathcal{M}_a $ followed by a post-selection $ \mathcal{P}_i=\ket{i}\bra{i} $ (also a CP map). Thus, it is natural to consider a three-time PDM such that there are two time intervals and a CP map takes place in each time interval.
We further assume that the two CP maps, $\mathcal{M}_a$ and $\mathcal{P}_i$, are implemented within the intervals $[t'_1,t'_2]$ and $[t'_2,t'_3]$, respectively. Denote the CJ matrix for the projector  $\mathcal{P}_i$ by $P^i$. According to Eq.~\eqref{eq: PDMclosedForm}, the three-time PDM for the two-time state  $ {}_{A_2} \bra{i} \otimes \ket{j}^{A_1}$ with Kraus operator $ \hat{E}_a $ is given by 
\begin{align}\label{eq:3timePDMai}
	R^{a,i}_{123} &= \frac{1}{2} ( R^{a}_{12} \, P^i_{23}  +   P^i_{23}   \, R^{a}_{12}), 
\end{align}
where $R_{12}^a = \frac{1 }{2} (M_{12}^a \ \rho + \rho \ M_{12}^a )$ with  $\rho = \ket{ j }_1 \bra{ j } \otimes \id_2 $. The probability to obtain the results $a$ and $i$ when no measurements are made at $t_1, t_2$, according to the probability rule in the PDM formalism, is given by (see Appendix~\ref{app: SimpPDMProb})
\begin{align}
	P_R(a,i)=\Tr R_{123}^{a,i} = |\bra{i}  \hat{E}_a \ket{ j }   | ^2 .
\end{align}
Then, the probability of obtaining the outcome $a$ given the post-selection is 
\begin{align}\label{eq: SimpPDMProb}
	P_R(a|i) =& \frac{ P_R(a,i)}{ \sum_{a'} P_R(a', i) } = \frac{ | \bra{i}  \hat{E}_a \ket{ j }   |^2 }{ \sum_{a'} | \bra{i}  \hat{E}_{a'} \ket{ j }   |^2 } ,
\end{align}
which is the same as in the two-time formalism given by Eq.~\eqref{eq: Simp2TimeProb}.

\subsection{PDM for the general pure two-time state}

We next consider a general pure two-time state   $ \sum_{ij} \alpha'_{ij} \, {}_{A_2}\bra{ i }  \otimes \ket{ j }^{A_1} $ with Kraus operator $ \hat{E}_a$. There are infinitely many ways to prepare this two-time state~\cite{aharonov2009multiple}. We emphasize that each realization of the two-time states can be assigned to a PDM. To simplify the investigation, we consider one particular realization while preserving the generalization of the same statistics obtained from this realization to any other realizations.
First, Alice prepares her system and an ancilla (denoted $C$) in the state
$$
\ket{\psi} = \sum_{ij} \alpha_{ij} \ket{j}_{A_1} \otimes \ket{i}_{C_1}\,\,  \text{ with } \alpha_{ij}= \alpha'_{ij} / \sqrt{\sum_{ij} |\alpha'_{ij} |^2}  ,
$$
at time $t_1$.  After performing her operation and keeping the ancilla undisturbed, Alice then post-selects the final state of her system and the ancilla to be the maximally entangled state 
$$
\ket{ \phi } = (\sqrt{d})^{-1} \sum_k \ket{k}_{A_2} \otimes \ket{k}_{C_2},
$$ where $d$ denotes the dimension of the system Hilbert space $\mathcal{H}_{A_2} $ and also that of the ancilla. Finally, trace over the ancilla, Alice's system is described by the two-time state. Alternatively, the general pure two-time state may be realized via Alice preparing the state $ \ket{ \tilde{\psi}} = \sum_{ij} \beta_{ij} \ket{j}_{A_1} \otimes \ket{i}_{C_1}$ at $t_1$ and post-selecting the state $  \ket{ \tilde{\phi} } = (\sqrt{d})^{-1}  \sum_k \gamma_k  \ket{k}_{A_2}  \otimes \ket{k}_{C_2}$ at $t_1$, with  $\beta_{ij} \gamma_i= \alpha_{ij} $.    

We now transfer to the PDM formalism. Once more, the physical realization can be modeled by a three-time PDM.
Denote the CJ matrices for the map $\mathcal{M}^a_A \otimes \id_C $ and  $\mathcal{P} _{\phi}:=   \ket{ \phi }  \bra{\phi}$ by $M^a$ and $P^{\phi}$, respectively. The three-time PDM for the two-time state  $  \sum_{ij} \alpha'_{ij} \, {}_{A_2} \bra{i} \otimes \ket{j}^{A_1}$ with the evolution $ \hat{E}_a $ is given by 
\begin{align}\label{eq:3timePDMaPhi}
	R^{a, \phi }_{A_1C_1A_2C_2A_3C_3} &= \frac{1}{2} ( R^{a}_{12} \, P^\phi_{23}  +   P^\phi_{23}   \, R^{a}_{12}), 
\end{align}
where $R_{12}^a = \frac{1 }{2} (M^a_{12} \ \rho + \rho \ M^a_{12} )$ with $\rho = \ket{ \psi }_1 \bra{ \psi } \otimes \id_2 $. Tracing over the ancilla $C$, the PDM for the general pure two-time state is obtained,
\begin{align} \label{eq: Gen2TStarPDM}
	R^{a, \phi}_{A_1A_2A_3} = \Tr_C R^{a, \phi}_{A_1C_1A_2C_2A_3C_3} .
\end{align}
The probability to obtain the results $a$ and $\phi$ while no measurements are made at $t_1,t_2,t_3$, according to the probability rule, is given by (see Appendix~\ref{app: GenTwoPDM})
\begin{align}
	P_R(a, \phi ) &=\Tr_A R^{a, \phi}_{A_1A_2A_3} =\frac{1}{d}  (\sum_{ij}   \alpha_{ij} E^a_{ij}  )^2  ,
\end{align}
where $  E^a_{ij}  := \bra{i} \hat{E}_a \ket{j} $.
Then, the probability of obtaining the outcome $a$ given the post-selection is written as
\begin{align}\label{eq: GenPDMProb}
	P_R(a| \phi )= \frac{ P_R(a,\phi)}{ \sum_{a'} P_R(a', \phi) } = \frac{ (\sum_{ij}   \alpha'_{ij} E^a_{ij}  )^2  }{ \sum_{a'} (\sum_{ij}   \alpha'_{ij} E^{a'}_{ij}  )^2  }, 
\end{align}
which is again the same as in the two-time formalism given by Eq.~\eqref{eq: Simp2TimeProb}.

\subsection{ PDM for the ensemble of two-time states }

Finally, we discuss the ensemble of two-time states $ \eta=\sum_r p_r \Psi_r \otimes  \Psi_r^\dag $ with Kraus density vector $J_a$. Here the density vector form is used, and the definition is given in Eq.~\eqref{eq: DensityVector}. 
In the previous subsection, we have assigned a three-time PDM to a general pure two-time state based on a specific experimental realization. Similarly, each pure two-time state $ \Psi_r \otimes  \Psi_r^\dag $ in the ensemble $\eta $ corresponds to a three-time PDM $ R_{123}^{a, \phi;r } $, taking the form of Eq.~\eqref{eq: Gen2TStarPDM}. In this way, the PDM for the two-time state ensemble $\eta$ is given by
\begin{align}
	R_{123}^{a, \phi } =  \sum_r p_r R_{123}^{a, \phi;r} .
\end{align}
The probability to obtain the outcome $a$ and $\phi$ with no measurements made at $t_1,t_2,t_3$  is given by 
\begin{align}
	P_R(a,  \phi )=& \Tr R_{123}^{a, \phi } = \Tr \sum_r p_r R_{123}^{a, \phi;r}  \nonumber\\
	=&   \frac{1}{d} \sum_r p_r (\sum_{ij}   \alpha_{r, ij} E^a_{ij}  )^2 .
\end{align}
The probability of Alice's operation given the post-selection is given by
\begin{align}\label{eq: ProPDMen}
	P_R(a|\phi)=& \frac{P_R(a, \phi)}{ \sum_{a'} P_R(a', \phi) }=\frac{\sum_r p_r (\sum_{ij}   \alpha'_{r, ij} E^a_{ij}  )^2 }{\sum_{ra'} p_r (\sum_{ij}   \alpha'_{r, ij} E^{a'}_{ij}  )^2 },
\end{align}
which gives the same probability as in the two-time formalism in Eq.~\eqref{eq: Prob2TimeEnsemb}.

In conclusion, this section has given a systematic method for assigning PDMs to two-time states with dynamic schemes based on their physical realizations. 
 We also emphasize that for coarse-grained CP maps $\mathcal{M}_a $, the probabilities in the above three scenarios have the same expression as Eqs.~\eqref{eq: SimpPDMProb}, \eqref{eq: GenPDMProb} and \eqref{eq: ProPDMen}, respectively.

\section{Mapping  bipartite process matrices with POVMs to PDMs }
\label{sec:PDMprocess}
In this section, we demonstrate how to assign PDMs to process matrices, along with a POVM, using two-time states as a bridge, as depicted in FIG.~\ref{fig:correspondence}. Previous investigation has shown that, for every process matrix, one can associate a two-time state that produces the same probabilities for all
POVMs~\cite{silva2017connecting}. We showed above that for every two-time state, one can find PDMs that produce the same statistics for all operations and now use that result to build a connection between PDMs and process matrices via the scheme of ~\cite{silva2017connecting}.

Given Alice and Bob's quantum operations represented as $\mathcal{M}_a$ and $\mathcal{M}_b$, or equivalently as Kraus density vectors $J_a:=E_a \otimes E_a^\dag $ and $J_b:= E_b \otimes E_b^\dag$, it has been shown that the bipartite two-time state~\cite{silva2017connecting}, 
\begin{align}\label{eq: Bi2TimeStat}
	\eta=\sum_{ \substack{ijkl\\ mnpq}} & w_{ijkl,mnpq}
	{}_{A_2B_2} \bra{ jl }  \otimes \ket{ik}^{A_1B_1} \nonumber\\
	& \otimes {}_{A^\dag_1 B^\dag_1} \bra{mp }  
	\otimes \ket{nq }^{A_2^\dag B^\dag_2},
\end{align}
produces the same statistics for Alice and Bob's operations as those in the process matrix 
\begin{align}\label{eq: BiProcessMatrix}
	W^{A_1A_2B_1B_2}=  \sum_{\substack{ijkl\\ mnpq}} w_{ijkl,mnpq} \ket{ijkl}\bra{mnpq },
\end{align}
that is,
\begin{align}
	\eta \cdot (J_a \otimes J_b)=\Tr W(M^a \otimes M^b)  .
\end{align}
In order to achieve our goal, it is necessary to link a PDM with the two-time state $\eta$ described in Eq.~\eqref{eq: Bi2TimeStat} in a manner that produces identical statistics for all POVMs.
As shown in Sec.~\ref{sec:PDM2TimeState},  one can construct such a PDM based on one physical realization of the two-time state  $\eta$. We state the explicit construction in the following.

For a bipartite two-time state 
\begin{align} \label{eq:eta}
	\tilde{\eta}=&\sum_{ r}{p_r } \Psi_r \otimes  \Psi_r^\dag \nonumber\\
	=& \sum_{ r}  \sum_{ \substack{ijkl \\ mnpq }}    p_r\,  \alpha'_{r,ijkl} \, \alpha^{'*}_{r,mnpq} \,
	{}_{A_2B_2} \bra{ jl }  \otimes \ket{ik}^{A_1B_1} \nonumber\\ 
	&\otimes {}_{A^\dag_1 B^\dag_1} \bra{mp }  
	\otimes \ket{nq }^{A_2^\dag B^\dag_2},
\end{align}
we take similar procedures to realize it as in Sec.~\ref{sec:PDM2TimeState}. 
It is worth emphasizing that there are infinitely many ways to prepare such a two-time state but here we only consider one way, as explained in Sec.~\ref{sec:PDM2TimeState} B.
For the pure two-time state $ \Psi_r \otimes \Psi_r^\dag $, an experimenter prepares the joint system for  Alice and Bob, and an ancilla (denoted $C$) in the state
\begin{align}\label{eq: InitBipartStat}
	\ket{\psi_r} = \sum_{ijkl} \alpha'_{r,ijkl} \ket{i,k}_{A_1B_1} \otimes \ket{j,l}_{C_1},
\end{align} at time $t_1$. Since $\Psi_r$ is a pre-selected quantum state, the normalization condition is not imposed here in order to simplify the probability calculations in the PDM and $\alpha'_{r,ijkl}$ can be any complex number.  After Alice and Bob perform their operations $\mathcal{M}_a \otimes \mathcal{M}_b$ and keeping the ancilla $C$ undisturbed, the experimenter then post-selects the final state of the joint system and the ancilla to be the maximally entangled state 
$$
\ket{ \phi } = d^{-1} \sum_{x,y} \ket{x,y}_{A_2,B_2} \otimes \ket{x,y}_{C_2},
$$ 
where $d$ denotes the dimension of Alice and Bob's Hilbert space $\mathcal{H}_{A_2} $ and $d^2$ is then the dimension of the ancilla. Trace over the ancilla $C$, the pure two-time state $\Psi_r \otimes \Psi_r^\dag$ is realized. Following the above procedure, but randomly preparing the initial state~\eqref{eq: InitBipartStat} according to the distribution $\{ p_r \}$,  the two-time state $\tilde{\eta}$ \eqref{eq:eta} is obtained.

Having the physical realization, the PDM can be constructed accordingly.
Each solution for the set of equations
\begin{align}\label{eq:SetEq}
	\sum_{ r}     p_r\,  \alpha'_{r,ijkl} \, \alpha^{'*}_{r,mnpq} &= w_{ijkl,mnpq},  \\
	\text{for}\,\, i,j,k,l,m,n,p,q&=0,...,d-1. \nonumber
\end{align}
identifies a PDM for the same statistics.

Take the process matrix $W^{A_1A_2B_1B_1} = \frac{1}{4} \sigma_0^{A_1} \otimes \sigma_0^{A_2} \otimes \sigma_0^{B_1} \otimes \sigma_0^{B_2} $ as an example to illustrate the map.  It is straightforward to verify that $W^{A_1A_2B_1B_1}$ is indeed a valid process matrix since it satisfies the two conditions proposed in Sec.~\ref{sec:pre} C.  The set of Eqs.~\eqref{eq:SetEq} for this process matrix is given by
\begin{align}
	\sum_{ r}    p_r\,  \alpha'_{r,ijkl} \, \alpha^{'*}_{r,mnpq} &= \frac{1}{4} \delta_{im}  \delta_{jn}\delta_{kp}\delta_{lq},  \\
	\text{for}\,\, i,j,k,l,m,n,p,q&=0,1. \nonumber
\end{align}
There are infinitely many solutions for the above equations, including 
\begin{align}
	\eta=\sum_{r=0}^{2^4-1} \frac{1}{2^4} \Psi_r \otimes \Psi_r^\dag,
\end{align}
where $ \Psi_r  = 2 \, {}_{A_2B _2} \bra{jl} \otimes \ket{ik}^{A_1B_1}  $ with $r=2^0 j+2^1l+2^2 i+2^3 k$. Given quantum operation $\mathcal{M}_a  \otimes \mathcal{M}_b$ with  corresponding CJ matrix $M^{ab}$, the 3-time PDM for each pure two-time state $\Psi_r$ is given by
\begin{align}\label{eq:3timePDMAi}
	R^{ab,jl;r}_{123} &= \frac{1}{2} ( R^{ab}_{12} \, P^{jl}_{23}  +   P^{jl}_{23}   \, R^{ab}_{12}), 
\end{align}
where $R_{12}^{ab} = \frac{1 }{2} (M_{12}^{ab} \ \rho + \rho \ M_{12}^{ab} )$ with  $\rho = 4 \ket{ ik }_1 \bra{ ik } \otimes \id_2 $, and $P^{jl}= \ket{jl}\bra{jl} $. Note that this three-time PDM is not normalized. Finally, the 3-time PDM for $\eta$ is given by $R= \sum^{15}_{r=0} \frac{1}{16} R^{ab,jl;r} $.

 While we found, in the above section, a mapping from PDMs to process matrices with a POVM, there are also important distinctions between the formalisms: (i) the PDM partly represents the POVM whereas the process matrix needs to be separately combined with the POVM elements to calculate probabilities, (ii) indefinite causal order in the process matrix formalism pertains to scenarios where events are delocalized in relation to a background time or because no such background exists, whereas in the PDM and multiple time state formalisms the events are measurements taken to be at well-defined times, (iii) to reproduce statistics in the process matrix formalism, post-selection is, at least in our mapping,  required in both the PDM and multiple-time state formalisms..

\section{Summary and outlook}
\label{sec:conclusion}
In this study, we investigated the relationship between three distinct methods of defining quantum states across different times and locations: the pseudo-density matrix (PDM), the process matrix, and the multiple-time state approaches. Previous research had demonstrated that bipartite two-time states could replicate the statistics of bipartite process matrices. We expanded on this by demonstrating that the operational scenarios underlying two-time states could be represented as PDMs, and we established a mapping from process matrices with a POVM to PDMs as a result.

The existence of this mapping implied that PDMs, like process matrices, could model processes with indefinite causal orders.  Our findings contributed to the integration of quantum models of spatiotemporal states, highlighting the potential of PDMs as a powerful tool for modeling quantum processes.

The results provide a systematic route for transferring results across the approaches. For example, it should be investigated whether the PDM negativity can be used to determine or partially determine causal structure in scenarios of superpositions of causal order.  We show in Appendix~\ref {sec:Switch} how the negativity of part of the PDM is intriguingly activated when two constant channels are placed in a superposition of orders. Moreover,
it would be interesting to study the relation between causal witnesses formulated in the process matrix approach~\cite{araujo2015witnessing} and the negativity of the PDM.

\vspace{0.5cm}

\noindent {\bf {\em Acknowledgements.}} We thank Fei Meng, Qian Chen, Zhida Zhang, Minjeong Song, Caslav Brukner, Vlatko Vedral and Dagomir Kaszlikowski for the discussions. We also thank the anonymous referee who helped us improve the manuscript. XL, FL and OD acknowledge support from the National Natural Science Foundation of China (Grants No. 12050410246, No.
1200509, No. 12050410245) and City University of Hong
Kong (Project No. 9610623). YQ is supported by the National
Research Foundation, Singapore and A*STAR under its CQT
Bridging Grant.
ZJ is supported by the National Research Foundation and the Ministry of Education in Singapore through the Tier 3 MOE2012-T3-1-009 Grant for Random numbers from quantum processes.

\section*{ Data availability  statement}

No new data were created or analysed in this study.

\section*{ Data availability  statement}

No new data were created or analysed in this study.

\appendix

\section{Derivation of Eq.~\eqref{eq: SimpPDMProb} } 
\label{app: SimpPDMProb}

In this appendix, we give detailed derivations of the probability as given in Eq.~\eqref{eq: SimpPDMProb}.

The CJ matrices $M^a, P^i$ is given by
\begin{align}\label{eq: CJMaPi}
	M^a  =& \sum_{mn } \ket{ n } \bra{ m } \otimes   \hat{E}_a  \ket{ m } \bra{n}   \hat{E}^\dag_a,  \nonumber\\
	P^i = &  \ket{ i } \bra{ i } \otimes    \ket{ i } \bra{i}  .
\end{align} 
Substituting Eq.~\eqref{eq: CJMaPi} into the PDM $R^a_{12}$, we have  
\begin{align}  
	\label{eq:2TimPDMa}
	R_{12}^a =& \frac{1 }{2} (M^a_{12} \ \rho + \rho \ M^a_{12} ) \nonumber\\
	=&  \frac{1}{2} \sum_{m,n }  \left[  \left( \ket{ n }_{1} \bra{ m } \otimes  \hat{E}_a \ket{ m }_2 \bra{n}  \hat{E}_a^\dag  \right) \ket{ j }_1 \bra{ j } \otimes \mathbb{1}_2 \nonumber \right. \nonumber\\
	& \left. +
	\ket{ j }_1 \bra{ j } \otimes \mathbb{1}_2 \,  \left( \ket{ n }_{1} \bra{ m } \otimes  \hat{E}_a  \ket{ m }_2 \bra{n}  \hat{E}^\dag_a \right) \right] \nonumber\\
	=&  \frac{1}{2} \sum_{m }  \left[  \left( \ket{ m }_{1} \bra{ j } \otimes  \hat{E}_a \ket{ j }_2 \bra{m}  \hat{E}_a^\dag  \right)  \right. \nonumber\\
	&\left. +
	\left( \ket{ j }_{1} \bra{ m } \otimes  \hat{E}_a  \ket{ m }_2 \bra{j}  \hat{E}^\dag_a \right) \right]  .
\end{align}
Next, combining Eqs.~\eqref{eq: CJMaPi}~\eqref{eq:2TimPDMa} with Eq.~\eqref{eq:3timePDMai}, we arrive at
\begin{align}
	R^{a,i}_{123} =& \frac{1}{2} ( R^{a}_{12} \, P^i_{23}  +   P^i_{23}   \, R^{a}_{12}) \nonumber\\
	=& \frac{1}{4} \sum_{m}  \Bigl( E^{a*}_{im}  \ket{ m }_{1} \bra{ j } \otimes  \hat{E}_a \ket{ j }_2  \bra{ i } \otimes    \ket{ i }_3 \bra{i}    \nonumber\\
	&+ E^a_{ij} \ket{ m }_{1} \bra{ j } \otimes     \ket{ i } _2 \bra{m}  \hat{E}_a^\dag \otimes    \ket{ i }_3\bra{i}  \nonumber\\
	&+ E^{a*}_{ij} \ket{ j }_{1} \bra{ m } \otimes  \hat{E}_a  \ket{ m }_2 \bra{ i } \otimes    \ket{ i }_3 \bra{i}   \nonumber\\
	&+  E^a_{im} \ket{ j }_{1} \bra{ m } \otimes  \ket{ i } _2 \bra{j}  \hat{E}^\dag_a  \otimes    \ket{ i }_3 \bra{i}  \Bigl),
\end{align}
where $ E^a_{ij} := \bra{i}  \hat{E}_a \ket{ j } $ and $ E^{a*}_{ij} := \bra{j}  \hat{E}^\dag_a \ket{ i } $. It is now obvious to see that the probability of obtaining the outcome $a$ and $i$ with no measurements made at $t_1, t_2$ is given by 
\begin{align}
	P_R(a,i) = \Tr R^{a,i}_{123}  = | \bra{i}  \hat{E}_a \ket{ j }  |^2.
\end{align}

\section{Derivation of Eq.~\eqref{eq: Gen2TStarPDM} }
\label{app: GenTwoPDM}

In this appendix, we give detailed derivations of the probability as given in Eq.~\eqref{eq: Gen2TStarPDM}.

The CJ matrices $M^a, P^\phi$ for the map $\mathcal{M}^a_A \otimes \id_C $ and the projector $\mathcal{P}_\phi= \ket{\phi} \bra{\phi}$, respectively, are given by
\begin{align}
	M^a =& \sum_{klpq } \ket{pq  }_{A_1C_1} \bra{ kl }  \otimes   \hat{E}_a \ket{ k }_{A_2} \bra{p} \hat{E}_a^\dag  \otimes \ket{ l }_{C_2} \bra{q}   , \nonumber\\
	P^\phi = &  \ket{ \phi }_{A_2C_2} \bra{ \phi } \otimes    \ket{ \phi } _{A_3C_3}\bra{ \phi }  \nonumber\\
	=& \frac{1}{d^2}\sum_{wxyz} \ket{ww} \bra{xx} \otimes \ket{yy} \bra{zz}   .
\end{align} 
The three-time PDM for the specific realization of the two-time state  $  \sum_{ij} \alpha_{ij} \, {}_{A_2} \bra{i} \otimes \ket{j}^{A_1}$ with dynamics $ \hat{E}_a $ is given by 
\begin{align}
	R^{a,\phi}_{A_1A_2A_3}=& \Tr_C R^{a,\phi}_{A_1C_1A_2C_2A_3C_3} \nonumber\\
	=& \Tr_C \frac{1}{2} ( R^{a}_{12} \, P^\phi_{23}  +   P^\phi_{23}   \, R^{a}_{12}), 
\end{align}
where $R_{12}^a = \frac{1 }{2} (M^a_{12} \ \rho + \rho \ M^a_{12} )$ with  $\rho = \ket{ \psi }_1 \bra{ \psi } \otimes \mathbb{1}_2 $ and $ \ket{\psi} = \sum_{ij} \alpha_{ij} \ket{j,i}_{A_1C_1}  $.

We proceed to calculate the probability to obtain $a$ and the state $\ket{ \phi }$ when no measurements are made at $t_1,t_2,t_3$,
\begin{align}
	&P_R(a, \phi) =\Tr R^{a,\phi}_{A_1A_2A_3} = \Tr_{AC} R^{a,\phi}_{123}\nonumber\\ 
	= &\Tr \frac{1}{4} ( M^a \, \rho \, P^\phi +\rho \, M^a \, P^\phi+P^\phi \, M^a \, \rho +P^\phi\, \rho \, M^a ).
\end{align}
There are four terms that contribute to the probability $P_R(a, \phi)$. Let us consider the first term first,
\begin{widetext}
	\begin{align}
		\Tr_{AC} M^a \, \rho \, P^\phi=& \Tr  \sum_{ \substack{klpq \\ ijmn} } 
		\left( \ket{p,q  }_{A_1C_1} \bra{ k,l }  \otimes   \hat{E}_a \ket{ k }_{A_2} \bra{p} \hat{E}_a^\dag  \otimes \ket{ l }_{C_2} \bra{q}  \otimes \id_3 \right)
		\left( \alpha_{ij} \alpha^*_{mn} \ket{j,i}_{A_1C_1} \bra{n,m }\otimes \id_2  \otimes \id_3 \right) \nonumber\\
		& \left(    \frac{1}{d^2}\sum_{wxyz} \id_1 \otimes  \ket{w,w}_{A_2C_2} \bra{x,x} \otimes \ket{y,y}_{A_3C_3} \bra{z,z}  \right) 
		\nonumber\\
		=&  \Tr_{AC}  \sum_{ \substack{p  ijmn  \\ wxyz} } 
		\Bigl( \frac{ \alpha_{ij} \alpha^*_{mn} }{d^2} E^{a*}_{wp} \ket{p,w  }_{A_1C_1} \bra{ n,m }  \otimes   \hat{E}_a \ket{ j }_{A_2}  \bra{x}  \otimes \ket{ i }_{C_2} \bra{x } \otimes \ket{y,y}_{A_3C_3} \bra{z,z}  \Bigl) \nonumber\\
		=&  \Tr_A  \sum_{ \substack{p  ijmn y} } 
		\Bigl( \frac{ \alpha_{ij} \alpha^*_{mn} }{d^2 } E^{a*}_{mp} \ket{p  }_{A_1} \bra{ n }  \otimes   \hat{E}_a \ket{ j }_{A_2}  \bra{x} \otimes \ket{y }_{A_3} \bra{y}    \Bigl) \nonumber\\
		=&     \sum_{  ijmn   }  \frac{ \alpha_{ij} \alpha^*_{mn} }{d } E^{a*}_{mn}     E^a_{ij}  = \frac{1}{d}  (\sum_{ij}   \alpha_{ij} E^a_{ij}  )^2
	\end{align}
\end{widetext}
where $E^{a}_{ij} :=  \bra{i} \hat{E}_a \ket{j}  $ and $E^{a*}_{mn} :=  \bra{ n } \hat{E}^\dag_a \ket{m  } $. Similar calculations for the other three terms yield the same result. Thus, we have 
\begin{align}
	P_R(a, \phi) = \frac{1}{d}  (\sum_{ij}   \alpha_{ij} E^a_{ij}  )^2 .
\end{align}

\section{ Activation of PDM negativity under quantum switch} 
\label{sec:Switch}

  The PDM negativity $f(R)$ introduced in Eq.~\eqref{eq: causmono}, which is forbidden by the other two formalisms, serves as a quantifier of quantum temporal correlation~\cite{fitzsimons2015quantum} and offers a causal perspective to investigating problems in the field of the foundation of quantum physics and quantum information. For instance, it has been employed to infer quantum causal structures~\cite{liu2023quantum}. Moreover, the logarithmic version of the PDM negativity has been found to be connected to the quantum capacity of a channel~\cite{Pisarczyk2019causal}.

In this Appendix, we, therefore, explore the PDM negativity under the quantum switch, an operation that superposes the orders of two consecutive channels. In quantum information theory, reset channels cannot transmit any classical information. People recently have found that if the orders of two consecutive constant channels are superposed and have access to the control system, information can be transmitted at a non-zero rate~\cite{Ebler2018enhanced,salek2018quantum}. This is known as information capacity activation. Considering the closed relation between information and thermodynamics, this phenomenon has also been explored in quantum thermodynamics~\cite{felce2020quantum,liu2022thermodynamics,nie2022,huan2022quantum}.
\begin{figure}[h]
	\centering
	\includegraphics[scale=0.9]{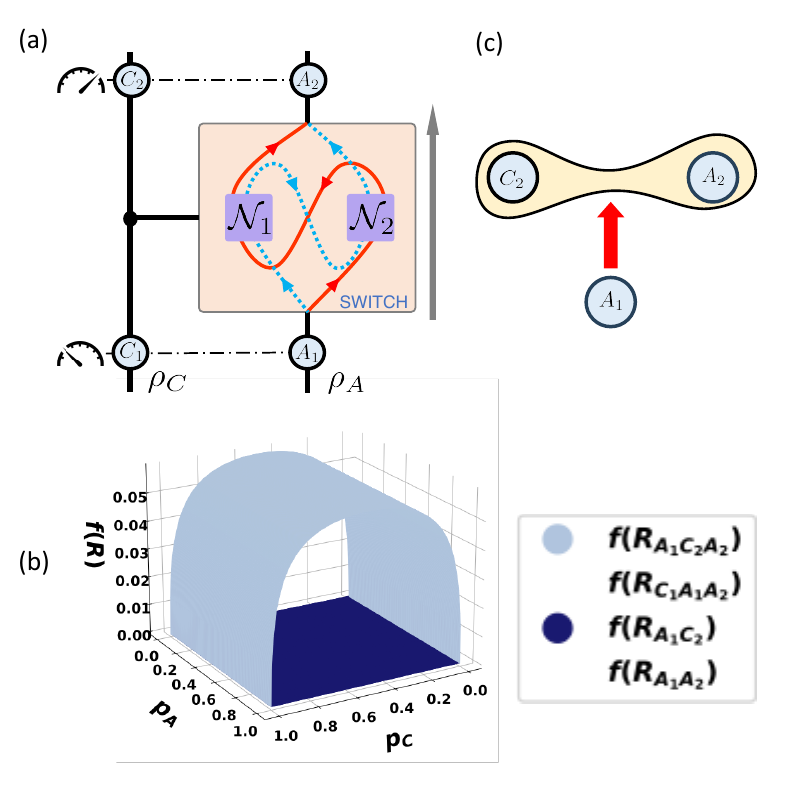}
	\caption{ {\bf{Quantum switch activates negativity in PDM.}} (a) The scheme for constructing the PDM of the quantum switch is depicted. The two orders of two constant channels,  $ \mathcal{N}_2 \circ \mathcal{N}_1$ and $ \mathcal{N}_1 \circ \mathcal{N}_2$, are controlled by the control system $C$. The coarse-grained measurements described by Eq.~\eqref{Eq:coarsegrainedM} are made at time $t_1, t_2$. (b) Simulation of the PDM negativities for $R_{A_1C_2A_2}, R_{C_1A_1A_2},R_{A_1C_2}$ and $R_{A_1A_2}.$  We observed that $R_{C_1A_1A_2}, R_{A_1C_2}$ and $R_{A_1A_2}$ are positive in the parameter regime, but $R_{A_1C_2A_2}$ can be negative.  (c) The causal structure of the three systems $A_1, C_2$, and $A_2$ is drawn based on the PDM negativities analysis. The `cloud' presents the correlation between $C_2$ and $A_2$. System $A_1$ has a causal influence on the compound system $C_2A_2$ but no influence on the individual systems $C_2$ and $A_2$.}
	\label{fig:switch}
\end{figure}

The setup, as shown in FIG.~\ref{fig:switch} (a), is taken from Ref.~\cite{liu2022thermodynamics} and is summarized as follows. A qubit $\rho_A$ goes into two consecutive uses of a constant qubit channel $\mathcal{N}$ and the orders of the two uses of $\mathcal{N}$ are controlled by the control system $\sigma_C$. In order to construct the total PDM $R_{C_1A_1C_2A_2}$,  coarse-grained measurements are made on the joint system $CA$ at times $t_1,t_2$. We assume that  initially, the joint state of control and system is in the state 
\begin{align}
	\sigma_{C_1} \otimes \rho_{A_1} = \ket{\psi_C}\bra{\psi_C} \otimes \ket{\psi_A}\bra{\psi_A},
\end{align}
where $\ket{\psi_{\beta}}:=\sqrt{p_\beta }\ket{0}+ \sqrt{1-p_\beta} \ket{1}, \beta=A,C $.
The constant channel $\mathcal{N}$ maps any state to the maximally mixed state, i.e., $\id/2$, and has the set of Kraus operators
\begin{align}
	\Bigl\{ &K_1= \sqrt{\frac{1}{2}} \ket{0}\bra{0}, K_2= \sqrt{\frac{1}{2} } \ket{1}\bra{0}, \nonumber\\
	&K_3= \sqrt{\frac{1}{2}} \ket{0}\bra{1}, K_4=\sqrt{\frac{1}{2}} \ket{1}\bra{1}   \Bigl\}.
\end{align}
Let us denote the Kraus operators of the channel $\mathcal{N}_1$ as $\{ K^{(1)}_i\}$ and $\mathcal{N}_2$ as $\{ K^{(2)}_i \}$. The resulting channel for superposing the orders of the two consecutive uses of $\mathcal{N}$ is expressed by 
\begin{align}
	\mathcal{S}_{\sigma_C}(\mathcal{N}_1,\mathcal{N}_2)(\rho_A) =\sum^4_{i,j=1}S_{ij} (\sigma_C \otimes \rho_A) S_{ij}^\dag, 
\end{align}
where $S_{ij}:=\ket{0}_c\bra{0} \otimes K_i^{(2)}K^{(1)}_j + \ket{1}_c\bra{1} \otimes K^{(1)}_jK^{(2)}_i$ denotes the Kraus operator for $   \mathcal{S}_{\sigma_c}(\mathcal{N}_1,\mathcal{N}_2)$. Denote $M$ by the CJ matrix of $    \mathcal{S}_{\sigma_c}(\mathcal{N}_1,\mathcal{N}_2)$, the constructed PDM for the joint system $CA$ across $t_1, t_2$ has the analytic form:
\begin{align}
	R_{C_1A_1C_2A_2}= \frac{1}{2} (M \, \rho + \rho \, M),
\end{align}
where $\rho=\sigma_{C_1} \otimes \rho_{A_1} \otimes \id_2  $.

By utilizing the PDM negativity $f(R)$,  we numerically study the information capacity activation phenomena from a causal perspective. It has been proven that for any PDM, the reduced matrix obtained by tracing out its subsystem at any time point is also a valid PDM~\cite{fitzsimons2015quantum,liu2023quantum}. This allows us to examine the temporal quantum correlations between subsystems within the corresponding reduced PDM. To study the information carried by $A_1$ passing on later systems, we calculate the negativity for the PDMs, $R_{A_1C_2A_2}, R_{C_1A_1A_2}, R_{A_1C_2}, R_{A_1A_2}$. Note that we have intentionally avoided $C_1, C_2$ appearing in those PDMs at once to eliminate the potential interference of the causal influence from $C_1$ to $C_2$.

From the simulation shown in FIG.~\ref{fig:switch} (b), we observe that 
$f(R_{A_1C_2A_2})>0$ for $p_0\in (0,1],p_1 \in [0,1] $ while $f(R_{C_1A_1A_2})=f(R_{A_1C_2})=f(R_{A_1A_2})=0$ for $p_0,p_1 \in [0,1]$. Although negativity in PDM is a sufficient but not necessary condition for temporal correlation, our simulation shows that there is causal influence from $A_1$ to $C_2A_2$, and there \textit{might} be no causal influences passing from $C_1A_1$ to $A_2$, from $A_1$ to $C_2$, and from $A_1$ to $A_2$. Therefore the simulation further implies that $A_1$ has a causal influence on the correlation within $C_2A_2$ but not on individual systems. The causal structure of the three systems $A_1,C_2$ and $A_2$ is drawn in FIG.~\ref{fig:switch} (c).  Finally, we note that the difference between $f(R_{A_1C_2A_2})$ and $f(R_{C_1A_1A_2})$ implies a time directionality, suggesting this approach may be useful to also analyse the arrow of time.

The negativity observed in those reduced PDMs might be understood as similar to mutual information in quantum information theory. In the quantum information setting~\cite{Ebler2018enhanced,liu2022thermodynamics}, there is a memory system $F$ that records $A_1$'s information and is kept undisturbed during the whole process. The information capacity activation refers to mutual information $ I(F: C_2A_2) \geq 0$, which corresponds to $f(R_{A_1C_2A_2}) \geq 0$. The quantity analogous to $f(R_{A_1A_2}),f(R_{C_1A_1A_2})$ is $I(F: A_2)$.  Mutual information $I(F: A_2)=0$ since the effective channel on $A$ is still a constant channel if $C_ 2$ is inaccessible. Finally, $I(F:C_2)$ is the corresponding information quantity for $f(R_{A_1C_2})$. It is shown~\cite{kristjansson2020resource} that information does not pass on to $C_2$, i.e., $I(F: C_2)=0$. Thus the PDM negativity allows us to infer whether the switch is ON as well as to infer the causal structure when it is ON or OFF. 

\bibliographystyle{apsrev4-1-title}
\bibliography{mybib}

\end{document}